Authors and affiliations:
(1) Deniz Marti, PhD
Harvard John A. Paulson School of Engineering and Applied Sciences (SEAS)
(2) Michael D. Smith, PhD
Harvard John A. Paulson School of Engineering and Applied Sciences (SEAS)

Motivating Data Science Students to Participate and Learn

ABSTRACT

Data science education is increasingly involving human subjects and societal issues such as privacy, ethics, and fairness. Data scientists need to be equipped with skills to tackle the complexities of the societal context surrounding their data science work. In this paper, we offer insights into how to structure our data science classes so that they motivate students to deeply engage with material about societal context and lean into the types of conversations that will produce long-lasting growth in critical thinking skills. In particular, we describe a novel assessment tool called participation portfolios, which is motivated by a framework that promotes student autonomy, self reflection, and the building of a learning community. We compare students' participation before and after implementing this assessment tool, and our results suggest that this tool increased student participation and helped them move towards course learning objectives.

## 1. Introduction

Data science programs are blossoming online and across university campuses. Besides courses heavily loaded with statistics, computer science, and other technical data science content, there is a growing recognition of the importance of courses that teach, as our colleagues from Berkeley have said, students to be "attentive to the social, cultural, and ethical contexts of the problems that they are formulating and aiming to solve" (Adhikari, DeNero, & Jordan, 2021). In other words, our educational approaches should encourage the students to treat data from and about humans with the same care that we expect in any other type of research involving human subjects.

In particular, data scientists need to be trained to understand current societal challenges, to listen carefully to the perspectives of the members of the communities from which they pull data, to productively critique each others' assumptions, and to communicate their ideas not only clearly, but with respect and beneficence for the individuals involved (NASEM, 2018). Numerous thought leaders have recently written about the importance of this set of skills (Chayes, 2021; Haas, Hero, & Lue, 2019; *Lue, R. A. 2019;* Wing, 2020; Irizarry, 2020; Adhikari, DeNero, and Jordan, 2021), and these leaders rightfully remind us, as David Madigan (2021) recently wrote, that the "range of data science and its impact on our daily lives raises challenging questions relating to privacy, ethics, and fairness."

We believe that the impact of data science on society will continue to expand and that we must regularly seek better ways to train the next generation of data scientists, especially in areas where there are no right answers, only better defended ones. What, then, are effective techniques we can use in our data science classes to develop in our students the non-technical skills that they desperately need to find success in their technical work? How can our courses help them to grow into stronger critical thinkers capable of considering the complexities of the human contexts surrounding their data science work?

This is not just a question of what content and methods to include in our curriculum, but a consideration of the intrinsic and extrinsic motivations at play in our assignments and classroom environments. Too often students approach the development of critical thinking skills with a focus on points and grades (a textbook example of extrinsic motivation). However, the intellectual and personal skills at the heart of critical thinking are more effectively developed by appealing to intrinsic motivations.

For example, we each bring our own perspectives, unique experiences, and biases to our work, and while the technical tools we use in data science may be applicable to a wide range of social circumstances (Ridgway, 2016; NASEM, 2018; He et al., 2019), it is often too easy for us to assume that our own perspective covers everything we need to consider in any data science problem. Courses like *Data 104: Human Contexts and Ethics of Data* at Berkeley (Adhikari, DeNero, and Jordan, 2021) and our own *Applied Computation 221: Critical Thinking in Data Science* (AC221) at Harvard present students with a selection of different perspectives through course readings, lectures, and exercises, but how do we get our students to make a habit of surfacing their underlying assumptions and confronting their implicit biases? How do we get them to engage deeply with the material and lean into the types of conversations and personal reflections that will produce meaningful and long-lasting growth in how they think?

This article describes how we have answered these questions in the context of AC221. We briefly cover the original organization of the class and how the transition to remote learning during the COVID-19 pandemic created a crucible in which we were forced to interrogate and question our own biases and blind spots about the standard pedagogy employed in these sorts of courses. The result was a new approach to participation and participation grading, which we argue is a crucial component in our classes to producing strong critical thinkers with the facilities to handle the complexities of human contexts in their data science work.

Our specific approach to participation and participation grading is backed by a general framework we call ARC, which integrates the theoretically-motivated pedagogical practices of autonomy, reflection, and community. Our experience is that this integration yielded deeper student engagement with the course's subject matter than we had seen in a previous instance of the course, and it created a foundation for students to receive feedback not only from the teaching staff, but also from each other. Overall, it helped create a community of learners who were motivated to collaborate, rather than to compete, with each other. And while we developed this framework for a discussion-heavy course like AC221, we believe ARC's utility goes beyond such courses; we are actively using it to create assignments and learning communities in a new, introductory programming class.

The paper is organized as follows: Section 2 briefly describes the goals and original structure of AC221, and it reviews what prior research considers best practices in student participation. It ends by recounting how the switch to an online learning environment in the spring of 2021 caused us to rethink our approach to student participation and participation grading. Section 3 presents our revamped approach, and Section 4 introduces a pedagogical framework to foster student motivation. This framework helps to explain many of the reasons for the success of our new approach, and we have found it to be a useful template for creating success in other types of courses. Section 5 provides an overview of our research methodology by which we test the hypothesis that students' participation improved in Spring 2021. Section 6 presentes the results where we compare across years of pre-intervention and post-intervention and indicators of success. Finally, Section 7 concludes with lessons learned and a look toward the future.

## 2. Traditional approaches to student participation

While AC221 is a fairly new course—it was first offered in 2018—the original pedagogical approach employed in the course reflected years of teaching case-based and discussion-heavy courses at Harvard. One of the authors (Smith) joined the course's teaching staff for the Spring 2020 offering, which was interrupted near the midpoint of the semester by the worldwide spread of COVID-19. With the cancellation of in-person instruction for the rest of that semester and Harvard's quick decision that all instruction for the entirety of the 2020-2021 academic year would be remote, we found ourselves in a unique position: From the forced transition in the Spring 2020 semester, we knew which aspects of the original, in-person course did not translate effectively to a remote learning environment. By knowing early in the summer of 2020 that we would have to teach remotely again in Spring 2021, we therefore had the time to address what had not worked well and adapt accordingly

## 2.1 Teaching critical thinking skills in data science

AC221 is a masters-level course, and its content highlights the wide-ranging impact data science has on the world. Its goal is to allow, encourage students to grow, foster student's growth in their ability to think critically about such thorny issues as fairness, privacy, ethics, and bias when they are building algorithms and predictive models that are then released into the world in the form of products, policy, and scientific research.

During the past three years, the course enrollment has varied between 40-70 students. For students within Harvard's data science master's program, it is a required class, but the course's focus regularly attracts graduate students from most of Harvard's professional schools. The result is a classroom of students with a wide variety of identities, experiences, and career goals.

The course's structure was specifically designed to take advantage of this diversity of student perspectives, and we actively encourage the students to interact not only with the teaching staff, but also with each other. For example, the foundational and case-based readings due prior to each class meeting are posted on Perusall[1], an online tool in which students collectively annotate the assigned readings. Ideally, Perusall allows discussions about the material to begin before the students gather together in class with the instructor. More importantly, a collective annotation system like Perusall allows each student to see what draws not only their interest, but the interests of the other students. This broadening of one's perspective is important because as data science's societal reach expands, so does the diversity of the stakeholders involved. Like the diversity of students in AC221, these stakeholders will exhibit a spectrum of different worldviews, biases, expectations, and interests.

Through the use of Perusall, along with class wide discussions, small-group breakout sessions, and a collection of collaborative assignments, our students learn to appreciate and handle a diversity of thought and perspective in the relatively safe environments of our classrooms before they find themselves grappling with it in the real world. In this way, the course exposes the students to the human elements associated with data science questions, and it provides them with opportunities to practice thinking critically about the real-world issues raised and thorny tradeoffs involved.

---

[1] https://perusall.com

## 2.2 Benefits and challenges in participation

The critical-thinking goals of AC221 are predicated on student participation that is genuine and effortful in the collective exercises and activities. In a learning environment where students fully participate, they develop the ability to generously listen to, question, and critique different perspectives. They begin to develop a sense of empathy and hone a healthy skepticism about what their data tells them about our world. In fact, research has shown that students' honest participation in discussion-based classes improves their critical thinking (Smith, 1977; Crone, 1997; Garside, 1996; Rocca, 2010), and that collaborative engagement with course material improves learning (Fritschner, 2000; Howard & Henney, 1998; Weaver & Qi, 2005).

However, instructors face significant challenges in accurately identifying the quantity and rigorously evaluating the quality of students' engagement (Dancer & Kamvounias, 2005; Rocca, 2010; Armstrong & Boud, 1983). For example, some types of participation may not reflect a student's actual engagement with the material. This is particularly problematic in assessment structures where participation is measured by simplistic rubrics, such as a count of how often students raise their hands or speak up in class (Armstrong 1978; Armstrong & Boud, 1983; Petress, 2006). While such counts are a measure of participation, they are often a poor proxy for understanding the depth of a student's engagement and growth.

Furthermore, such poorly designed structures provide no encouragement for students to communicate their authentic opinion. It can be too easy for students to play it safe, to parrot what they think the instructor wants to hear, and to agree with the dominant opinions. These tactics satisfy simplistic participation rubrics, but they inhibit what we wish to achieve: the illumination of hidden assumptions and implicit biases that lead to personal and intellectual growth. They do not encourage students to take chances, make mistakes, and learn from them.

Even in instances when instructors are able to capture the quantity and quality of student participation, they are further challenged by the time and effort in providing the students with timely and actionable feedback on their participation performance. It is not uncommon in many well-structured engineering, business, and law classes to provide feedback on participation performance through a letter grade distributed only a few times during the semester. In contrast, research has shown that timely feedback that targets a specific skill is crucial for learning.

Ideally, learners develop mastery through a specific type of practice, called *deliberate practice*, which involves performing a particular skill in a context under guidance, receiving immediate feedback on the performance, and then being given an opportunity to incorporate that feedback into subsequent practice (Ericsson, Krampe, & Römer, 1993). This gives the learners the opportunity to authentically reflect—a key component in learning—and is crucial for honing one's ability to apply the knowledge in real-life contexts (Herrington & Oliver, 2000). Rote practice, on the other hand, is typically decontextualized and lacks productive feedback or coaching. Without timely and actionable feedback, practice becomes a suboptimal process for learning because it is too easy for mistakes to go unnoticed, be repeated, and become ingrained in our brains. And without reflection, practice becomes inefficient (i.e., we waste time focusing on what we already know rather than on those areas where we need to improve) and the individual gains remain ephemeral. Specifically, reflection encourages learners to create connections between their separate learning experiences.

As a final challenge, instructors must decide upon an overarching approach to assessing and grading participation. As discussed above, the approach should build in timely and actionable feedback on a student's participation and include ample opportunities for the students to reflect on their individual progress. Equally importantly, the grades given for participation should focus on each student's efforts rather than their performance relative to their peers. Participation is meant to put students on a path to engagement with the course material and each other. The grading of participation should not dissuade risk taking, which can accelerate learning and growth. In general, the overarching approach to participation should create an environment for personal and collective learning. It should avoid any feeling of competition. While the setting of minimum participation expectations can help launch students on this path, assessment of participation should be largely formative in nature and not strictly summative.

## 2.3 Participation grading pre-pandemic

When AC221 began in Spring 2020, we approached participation as a necessary and important but largely unremarkable part of the class. Looking back at our syllabus from that year, we mention the word "participation" only once, when describing the weights given to the different portions of student work in calculating their final course grade.

We did take time in the first class meeting, as we had done every year, to describe how the students could participate in our class. Specifically, we outlined the venues in which students could demonstrate their participation, namely in-class questions and discussions, during the small-group breakouts, by leaving comments in Perusall, and contributing items in the "Current Events" portion at the start of each class. We also covered our expectations, both for how the students would engage with each other to create a safe learning environment for all and what we as instructors would consider to be a satisfactory level of participation.

In hindsight, it's likely that what we envisioned for participation was never clear in the students' minds. We regularly received a few questions about it in the first class, but additional questions did not arise until the semester's mid-point when we finally provided the students with their first formal feedback. Except for the few students who were obviously disengaged, most students did not receive any other formal participation feedback until we determined their final grades.

Of course, the students did at times receive some feedback from us about their participation. We might follow an in-class student comment with a few words of praise if the point was obviously thoughtful, or sometimes offer thanks if a student was obviously trying hard to contribute. And through the use of Perusall, we were able to give students who found it hard to participate in the moment alternative avenues to contribute their thoughts and perspectives. Perusall, for example, allows instructors to 'upvote' student comments with a single click and insert their own comments into an ongoing thread. Like our words of praise and thanks in class, this feedback helps everyone in the class identify what the instructors think is good participation, but the public nature of these channels are not ideal for giving the full spectrum of rich feedback we would like to sometimes give individual students.

Overall, we knew that our approach to encouraging, reacting to, and assessing student participation was imperfect. At times, we would begin a class reminding the students that a Perusall or in-class comment consisting of nothing more than "I agree with Sergey" or "Great thought, Sue!" was not what we hoped to see in our collective discussions. We were also aware that some students felt like they needed to constantly participate no matter how much the topic at hand interested them. This feeling worked directly against our desire to have the students participate with their authentic selves. Instead, the students tended to take relatively safe stances in our discussions.

And *safe* was a good way to describe how we as instructors chose to weigh participation in the overall grading of the course. Pre-pandemic, participation contributed just 10 percent of a student's final grade in AC221. This was despite the fact that the weights on the different components of a course's final grade are a significant signal to the students; the students interpret these weights as saying where the instructor thinks the students should spend their time and energy.

Something smelled rotten in our approach, but the smell wasn't so strong (or we were sufficiently acclimated to it) that we had felt a strong need to clean up the mess. Plus, despite these imperfections, the course's content and the regular appearance of this content in the daily news often ignited some phenomenal conversations. In those times, the students lowered their academic shields, learned from each other, and truly grew as critical thinkers.

## 2.4 The move to remote learning

Then came the global pandemic and the ceasing of in-person instruction. Discussions that might have gained energy in the classroom often fizzled out in Zoom. Topics that used to engage the students in class and continue as they filed out of the classroom at the end of our time together stopped dead as students blinked out at the end of our Zoom session.

In a physical classroom, experienced instructors have learned to 'read the room' and get a sense of whether the students are actually engaged, and when not, these instructors have their personal toolbox of little things that they use to reignite discussion. Experienced instructors have also learned to constantly shift their focus around the physical classroom as it is hard for a student to stay disengaged when a student nearby is drawn into the conversation. While Zoom and its equivalents put every student in the front row, these tools do not recreate the power of proximity we find in a physical classroom.

We do not mean to say that a remote learning environment is categorically worse than a traditional classroom environment. We do not believe that is true. Our point is that they're different, and these differences matter in how we organize and teach our students. If we had any doubt, the stark difference between the beginning and end of the Spring 2020 offering of AC221 made this abundantly clear to us. It showed us that we could no longer rely on physical proximity in our classrooms to make up for the imperfections in our approach to participation. Fixing this to achieve the level of student engagement and interaction the course's learning goals demanded instantly became our top priority.

Diving into this work, we quickly realized that this was not a problem that could be solved in its entirety using existing technology solutions that help instructors be more rigorous in their participation grading and more inclusive in their attempts to draw the students into the discussions. These thoughtfully designed tools made improvements to the traditional approach to participation and participation grading (i.e., the approach we described in prior subsection), but the end of the Spring 2020 semester made clear that we needed something more: we needed to fundamentally rethink our approach.

## 3. A new pedagogical approach: participation highlights

Our first design decision was to boost participation from 10 percent of the final grade to one-third of it. Participation was now equal to the total weight of the eight short (two-page) critical-thinking papers that the students had to write during the semester. To make the final grade calculation work numerically, we took weight from the three programming assignments and the student's final project, making each of those categories one-half the weight we were assigning to participation.

It's not the specific numbers here that matter, but the relative weightings. We now felt like we were making a statement we could not ignore and the students could not miss. To execute on this consequential choice, we found that we had to answer a number of fundamental questions.

## 3.1 What qualifies as participation?

To make any of this work, we and the students needed a shared understanding of how the students could participate in the course. It was not too hard to start this list, and we took advantage of the students' desire to know *how* to simultaneously slip in an explanation for *why* participation was important. Specifically, here's the opening of our syllabus section titled *Participation Grading*:

> *To deepen your engagement with this course's material, which will help you better learn it, a portion of your final grade depends upon your participation in the parts of this course that involve us learning together, as a community of learners. We refer to the following parts as the collaborative-learning parts of AC221:*
>
> *1. Time spent collaboratively annotating the Perusall readings*
> *2. Class time spent discussing current events*
> *3. Instructor questions posed to the entire class*
> *4. Student questions asked while we're together as a class*
> *5. Time spent in breakout groups*

It did not take long for us to learn that this was not an exhaustive list. As we detail below, we encouraged the students to take ownership of their participation, and to our delight, they even did so in defining what qualifies as participation. For example, the students made great use of the Zoom chat feature to extend and enrich our live discussions. Normally, instructors lament the encroachment of electronic channels of communication in our classrooms, as they feel that these distract the students' attention. All it took to flip chat from a distraction to a learning tool was our agreement that the students could use Zoom chat as a way to demonstrate good participation.

**3.2 What is quality participation?**

But you might ask and the students definitely wondered, what qualifies as *good* participation? To answer this question, we began with a clear statement of what we would *not* be tracking and grading. In this regard, we emphasized two points: (1) We would not be taking attendance. The students were expected to attend our discussion-based class, and we reminded them that they could not perform well in participation if they were not regularly there to participate. And (2) we would not be counting the number of times that they spoke up. We told them that what matters in assessing their participation is the *quality* of their comments and questions, not the *quantity*.

Quality, we explained, stems from the impact of their participation on the growth experienced by themselves and their classmates. In a qualitative study that analyzed the characteristics of participation activities, students' efforts to incorporate ideas and experiences were found to be one of the most significant strategies that increase quality participation (Dallimore, Hertenstein & Platt, 2004).

Prior work suggests that relying on *quantity*, such as taking attendance or roughly counting raised hands may be ineffective or misleading when assessing students' participation (Armstrong & Boud, 2006). Therefore, it is important to identify the cues that indicate quality participation and incorporate those cues into an assessment tool that measures participation reliably and rigorously (Danser & Kamvouias, 2010). While prior research has attempted to operationally define participation, it fails to capture the underlying mechanism in which quality participation is indicated. Fassinger (1995) suggests that commenting on a topic and asking a question are the indicators that qualify students' activity as participation. This categorization captures the participation phenomenon at a superficial level, thus it fails to operationally define the underlying mechanism of a quality participation. For example, according to such operationalization of participation assessment, student comments such as "I agree with Susie" or questions such as "Can you explain this topic?" would be considered "participation." However, we argue that such comments or questions do not necessarily qualify as high-quality participation because they do not demonstrate a level of elaboration (e.g., building on an argument) or extension of the discussion (e.g, providing a counter-argument to an existing one). Therefore, we argue that quality participation is not simply agreeing or disagreeing with the instructor or a classmate.

In this course, AC221, we encourage and value students' activities such as critiquing the ideas of others and communicating their own ideas with evidence and clear reasoning. Students are expected to demonstrate understanding through the raising in our collective discussions of hidden connections across the course material. Asking questions that spark discussion was considered as important as providing answers or adding comments that invite further discussion.

## 3.3 How often to participate?

We found that it was equally important to address the issue of quantity. On the syllabus, we said:

*"We'd rather have you speak up, for example, every other class period (or on 40% of the readings) and say something interesting each time than have you speak up multiple times per class with comments or questions that don't push forward the conversation".*

Unsurprisingly, students are not equally interested in every topic, question, or paper we cover. If a student believes that she must demonstrate participation in every aspect of the course, this grading concern diverts part of her attention and interferes with the type of engagement that leads to learning. Even if the student is intrinsically interested in the current question, we often saw students more focused on the issue of participation than their interest in the topic.

We chose to push the students' natural concern for the grading of participation from something that was constantly in their mind on every aspect of the course to one that they had to visit only intermittently. In particular, we told the students that we would require from them only a few instances of participation over several course meetings. This immediately removed the pressure that they needed to find some, often unnatural, way to participate in every activity.

Furthermore, this approach created a perfect alignment between our concern as instructors (i.e., that the students would check out of the class for long periods of time) and the students' concern about participation grading (i.e., that they had not shown any engagement with the course over the last few activities). From a pedagogical point of view, this approach caused the students to practice spacing (Kang, 2016) while also giving them the autonomy to choose when and how to engage with the material.

## 3.4 Participation portfolios and highlights

Once we had given students ownership over their participation and how we would grade it, we wondered if giving even more agency to the students might solve other historical teaching challenges. In particular, as discussed earlier, participation grading is difficult for instructors because a student's observable actions do not always provide the information instructors need to rigorously evaluate the quality of that student's engagement with the material. To avoid this need to read minds, we realized we simply had to ask the students to explain what was in their minds. Specifically, we asked the students to create and maintain a participation portfolio in which they would document evidence of their own *good* participation across the course. The form of this portfolio could be anything they wished as long as it was a place they found convenient to compile, over time, selections of their participation efforts and from which they could later choose the best of these efforts.

To be clear, we never asked the students for their participation portfolios. Instead, we asked them to submit *participation highlights* drawn from this portfolio. Specifically, the AC221 syllabus for Spring 2021 stated that we were breaking the semester into 7 participation periods (of a length of typically three class meetings), and that at the end of each of these periods, we wanted the students to reflect and submit what they believed to be their three best examples of participation.

We did not, however, require three good examples for the students to receive full credit for the participation period. We told the students that only two of these three highlights had to be good examples. We asked for three, we said, so that the student could take a chance with their third one on something new and receive feedback from us without penalty. This was one of our ways of making concrete what we regularly told the students: *"You don't learn as fast if you aren't making mistakes and learning from them."*

Honestly, two-things-over-three-classes may sound like a low bar, but we found it to be more than many students had historically done in this course. In the past, a student could check out for weeks on end, but under this new approach, we received a number of student complaints that they were always working in the class!

The careful reader will notice that we have not actually solved the problem of instructors having to read the minds of the students. To this point in our description, we have simply removed the need for the instructor to note and record each student's participation.

We solved the mind-reading problem in what we required the students to submit in each participation highlight. In particular, it was not sufficient for the students to just tell us how they participated at some point in the class, but we required them to also reflect for us on why that instance of participation demonstrated their own intellectual growth or contributed to the intellectual development of some of their peers. And in this reflection, we asked for evidence. The syllabus said,

*"You can help each other to generate examples for your participation portfolio. For example, if you have an honest question about some part of a Perusall reading, a topic in lecture, or a point*

*in a breakout session, ask it. Then if another student provides you with a particularly illuminating*
*answer to your questions, take a moment and send that peer a note expressing how you found*
*their answer helpful. [This] exchange [is] a good example of participation."*

Asking the students to submit their participation highlights after every three (or so) classes meant that we could no longer provide a single, mid-semester note to the students about their efforts in participation. As instructors, we would have to grade and provide feedback now seven times (for this instance of AC221). While this was a real cost borne by the instructors, we found the work enjoyable. We now were reacting directly to circumstances and issues raised by individual students. We were able to comment equally on good and bad participation. We felt we were finally helping all the students.

And while authentic practice depends upon the students receiving immediate feedback on their performance and the opportunity to incorporate that feedback into subsequent practice, we learned that it was not just our regular feedback that the students incorporated into their subsequent practice. This approach allowed the students to provide each other with real-time feedback.

## 4. The ARC framework

Our approach to participation and participation grading draws upon insights from the cognitive psychology and education literature, a number of which we discussed earlier. At the heart of the approach is a set of three individual characteristics well known in this literature—autonomy, reflection, and community—that we link in mutually-reinforcing ways. Considering them together, we are able to create engaging and effective assessments. We call this framework ARC, which stands for "Autonomy-Reflection-Community" (see Figure 1), for it links three important characteristics that are essential for creating an environment where students share their ideas in a community of learners and strengthen their learning through reflection.

While we assess the students individually in activities built with ARC, it is important to understand that we are not talking about solitary activities, but activities done together as a group, as a community of learners. The sharing of perspectives, such as occurs in a class discussion, or the explicit interactions between classmates, such as are found in team-based learning approaches, are two examples of often-employed group activities. Participation in AC221 is simply another example of an assessment of individuals done during a group activity. And all of these examples are activities that the instructor of a course has identified as crucial to achieving that course's learning objectives. In summary, ARC is an approach for helping instructors build effective assessments for discussion-based classes in which students' participation plays a vital role.

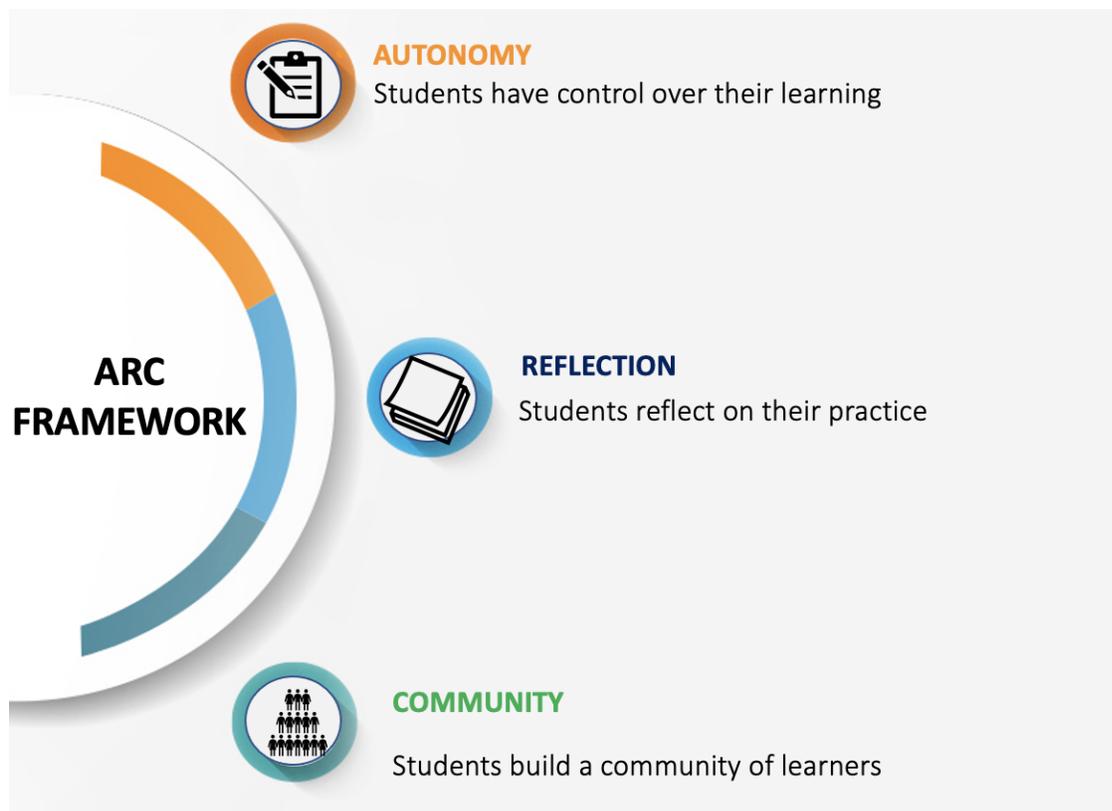

Figure 1. The ARC framework for assessment design.

Let's now see how ARC structures group activities to foster student engagement and enhance student learning using AC221's approach to participation as a running example. Overall, an assessment built within the ARC framework encourages each student to: (1) take ownership of their learning through an actionable level of self-direction or autonomy in the group activity; (2) reflect on their personally selected practice during and after this activity; and (3) build on peers' perspectives by following up with previously stated opinions during the group's activity.

## 4.1 Autonomy

A student's introduction to an ARC-based assessment begins with a dose of autonomy. Education psychologists posited that the sense of autonomy improves students' intrinsic motivation, engagement with activities, and in turn their willingness to learn (e.g., Benware & Deci, 1984; Grolnick & Ryan, 1987). Therefore, it is critical to design assignments in a way that they enhance students' sense of autonomy (e.g.., November, 2012). Stefanou et al., (2013) explored perceived autonomy in courses that require higher-level thinking  and concluded that students reported the autonomy support that they received from the instructor led them to be independent thinkers. Research shows that course design interventions that foster autonomy led students perceive more ownership over their learning, leading to numerous positive outcomes such as persistence and educational achievements (Vansteenkiste et al., 2004; Guay & Vallerand, 1996;Yu & Bristol, 2020; Bao & Lam, 2008). Boud (2001) argued that students' autonomy is crucial for proactive learning, leading to greater responsibility and agency, in contrast with reactive learning, a form of attitude where students simply react to the stimuli provided by the teacher.

In AC221, this was accomplished by giving the students the opportunity to choose from among the many ways in which they could participate in the course's collective activities. This was easy for the students to understand (i.e., it was actionable), and as such, it began to shift participation from something decreed by the instructor into something that could be owned by the students. The students were then given further autonomy by being able to decide which instances of participation (as long as they had attempted more than two in the past several class periods) they wanted the instructor to grade as their participation. Students could base these decisions on:

- their interests (e.g., one paper might interest them more than another);
- their level of comfort with particular modes of participation (e.g., if they are uncomfortable raising their hand in class then they can choose to participate asynchronously through Perusall); or
- their self-assessment of which instance of participation best exemplifies their performance (e.g., in the selection of which highlights they submit).

As long as the course design provides a sufficient variety of pedagogically similar venues for participation, this freedom of choice gives the students a sense of control and ownership over their learning. Although these examples are specific to AC221, autonomy can be fostered in any class through the development of course-appropriate mechanisms. As long as the course design provides a sufficient variety of pedagogically similar venues for participation, this freedom of choice gives the students a sense of control and ownership over their learning.

**4.2 Reflection**

Once a student has made a choice and begun to engage in the group activity, an ARC-based assessment next encourages the student to observe and reflect on their performance, both during and after it. Research suggests that self-regulated learning is supported by the processes of self-observation and self-judgment (Zimmerman, 1989). Self-observation is a process in which students systematically monitor their own learning, while self-judgment requires students to evaluate their activity based on the desired learning goal (Zimmerman, 1989). Engagement and participation were shown to increase in course designs where self-observation and self-judgment activities were implemented (Zaremba & Dunn, 2004; Delprato, 1977). Enhancement in self-regulation learning influences self-efficacy, i.e., confidence in one's ability to learn or develop a skill in a given context (Bandura, 1977). In turn, greater self-efficacy has been shown to enhance learning and motivation (Deci & Ryan, 2000; Caprara et al., 2008). Therefore, it is important to include a reflective aspect when designing an assignment.

At the start of AC221, we made the students aware that they needed to capture and record their instances of participation, and we repeatedly reminded them of this in a hopefully memorable way through our regular reference to their participation portfolios. The students, therefore, began each of the course's group activities in a mode of self-observation. Furthermore, as we (the instructors) and their peers reacted in the moment to a student's participation, this student was primed to reflect: Was that the type of reaction they expected from their comment or question, and is it an instance that they should include in their participation portfolio?

Students were not only primed for reflection in individual instances, but this behavior also became ingrained. Given the frequent, relatively low-stakes nature of the participation highlight assignments, self-observation and self-judgment became a regular part of the students' approach to the class. In addition, autonomy encourages authentic participation, allowing the students to reflect on their learning journey rather than simply report on their forms of participation.

Finally, the reflection is spaced. Students reflect in the moment. They reflect as they record instances in their participation portfolio. They reflect again as they decide which recorded instances to submit as their participation highlights every other week. And they reflect once more as they receive the instructor feedback on their highlights.

**4.3 Community**

We include community in ARC not only because it describes the kind of assessments for which the framework succeeds, but also because we find that encouraging the students to undertake a group activity *as a community of learners* deepens and enriches the student reflection that takes place during it. As we said at the start of this section, the ARC framework encourages the students to share informal feedback with their peers as they work together.

Ramsden (1992) showed that students learn better in discussion groups, compared to simply listening to the instructor without any peer interaction. Other research suggests that students with a greater sense of relatedness, or as Deci and Ryan (2000) describe, "integration of oneself within the social community," are more likely to foster engagement. In his seminal work, Watkins (2005) argued that there are certain hallmarks of a community of learners: owning agency, developing a sense of belonging, improving cohesion among the learners, and welcoming diversity of opinions.

Beyond these insights from the literature, we use a learning community to address the challenge that instructors cannot always provide timely feedback on every action by a student in a group activity. If the students themselves understand the importance of peer feedback and can benefit personally from participating in it (i.e., peer feedback doesn't come with negative externalities), peer feedback can encourage reflection in the time period until an instructor's feedback is available.

In AC221, we explained that students could help each other recognize good examples of participation by sending each other notes in which they expressed when and why a peer's participation helped to broaden their perspectives or advance their understanding of the material. The student receiving this note could then use this informal feedback as evidence of the impact of their participation not only on their own learning, but the learning of the larger community. This then fed upon itself as the students began to seek out the types of participation that would lead to meaningful discussions. As a result, students began to see value in listening to each other, challenging each others' assumptions and views, and answering each others' questions.

**5. Research Methodology**

We hypothesized that our participation assessment approach built upon the ARC framework, which draws its inspiration from the existing literature, improves the quality of students' participation. To test the effectiveness of this intervention, we compared student participation in the Spring 2020 version of AC221 (pre-intervention) with that in Spring 2021 (post-intervention). In what follows, we describe the data we gathered, its structure, and how we came to the metrics we used in assessing the quality of student participation in this data. We then present the specifics of the rubric we used in scoring the data.

## 5.1 Data gathered

We considered the different aspects of how the students can participate in AC221 and chose to focus our data gathering and data analysis on the students' Perusall postings in the two years. This choice supported our hypothesis testing in several ways:

1. The majority of Perusall assignments were the same in both years, giving us a large and consistent baseline for comparison. Each year included other Perusall assignments that were unique to that year, and we removed those from the data set we built.

2. Every student posting on Perusall is permanently recorded. We have exactly what each student said in a discussion and the context in which they said it. This promotes an unrushed, direct, and objective evaluation of the student's participation as it avoids the need for any on-the-fly rating by an instructor or a third-party, and any self-rating assessment of participation by the students themselves, as has been done in other research evaluations of participation (e.g., Frymier & Hauser, 2016).

3. As a venue for participation in AC221, Perusall was the one that didn't change from the mix of in-person and online instruction in 2020 to solely online instruction in 2021. We were particularly concerned with the risk of confounding factors arising in the noticeably different modalities of "classroom" discussions and the way we handled breakout groups between the two studied years.

We assert that the Perusall environment is a good proxy for the in-class discussions as it mimics the classroom where students can comment, ask questions, and discuss materials with one another. In fact, the instructors encouraged the students in both years to consider this online environment as an extension of the classroom.

Perusall is an online platform for students to collaborate on their assigned readings. Students highlight a passage or an area of a figure in a reading and annotate that passage/area with a *comment*.[2] This comment might consist of one or more statements or questions. Other students see that comment and can upvote and/or reply to it. The initial comment starts what Perusall calls a *conversation*. Comments by other students in reply to the initial comment, including further comments by the initiating student, are displayed in chronological order in Perusall's conversation pane. Students are capable of upvoting any of the comments in a conversation.

---

[2] https://support.perusall.com/hc/en-us/articles/360033995074-Getting-started).

In Perusall, the highlighting of two different passages or areas creates two different conversations. Two highlighted areas that overlap but are not exactly the same also create two different conversations.

## 5.2 Conversations vs. discussions

Our analysis defines separate metrics for Perusall comments and conversations. However, we refer to Perusall conversations as *discussions* since that is what we hope them to be in reference to what students think about their participation. Most people think of a conversation as any exchange between individuals, while a discussion is reserved for conversations about a specific topic or toward a specific goal (Michigan Publishing, 2008).

## 5.3 Measuring participation through comments and discussions

To measure changes in student participation reliably and rigorously, we base our metrics on cues that indicate progress toward the learning goals in AC221. In discussion-based classes like AC221, it is crucial for students to acquire and use skills such as articulating their authentic ideas, being open to differing opinions, and demonstrating the integration of multiple perspectives around the class material. Students more quickly become independent thinkers as well as contributors to the community by sharing their authentic opinions in discussions with others, instead of passively sitting and absorbing the instructors' "pre-packed knowledge" (Cacciamani et al. 2012). As such, we assess student participation through:

1. how well the students include authentic opinions in the construction of their Perusall comments; and
2. how much a discussion makes sense of and locates peers' points of views, and how well it synthesizes a wide range of diverse opinions into the overall dialogue.

**Authenticity of comments.** To rigorously identify the factors that indicate sharing authentic opinions in a student comment, we relied heavily on the work of Hadjioannou (2003; 2007). It states that expressing one's own ideas as well as reflecting and connecting the subject matter or peers' arguments with one's own experience are the determining factors for authenticity (Hadjioannou, 2003). Therefore, we assess authenticity as the extent to which students share personal examples, opinions, experiences that are relevant to the class subject. Students are expected to elaborate on the connection between the subject matter and their authentic expressions, rather than simply sharing a personal example.

**Quality of discussions.** In discussion-based classes, authentic comments are only as important as their contribution to the quality of a discussion. A high-quality discussion is where students build upon each other's ideas while constructing their own perspectives. Our metric for measuring the quality of a discussion, therefore, relies heavily on prior work in which the stages of perspective taking in an asynchronous learning environment are shown to correspond to the quality of the discussion (Häkkinen & Järvelä, 2006; Järvelä & Häkkinen, 2002).

Drawing upon seminal work about social cognitive development models, Häkkinen & Järvelä (2006; Häkkinen, Järvelä, & Mäkitalo, 2003) posited that discussions are low-quality when students hold egocentric opinions and fail to acknowledge that others may possess different perspectives (Selman, 1971; Greeno, 1998; Häkkinen, Järvelä, & Mäkitalo, 2003; Häkkinen & Järvelä, 2006). Due to this lack of comprehension and appreciation of diverse perspectives, the discussions typically do not advance, and the replies tend not to respond to or take into consideration previously stated ideas (Häkkinen & Järvelä, 2006).

On the other hand, a high-quality discussion is one in which students are clearly confronting their implicit biases, building upon their peers' ideas, and including a rich set of cross-references. Discussions that "recognize and value the uniqueness of each person's opinions and expressions" (Järvelä, Häkkinen, & Oostendorp, 2003) are high quality because they advance the learning objectives of the course. This includes understanding modern social contexts through various lenses, the values and opinions of the communities from which data is drawn, and the ability to communicate about diverse perspectives respectfully. Furthermore, these discussions advance when students provide counterarguments, challenge assumptions, and acknowledge peers' perspectives (Schaeffer et al., 2002).

**Caveats.** These metrics are not meant to capture every type of response or exchange that promotes learning in a discussion-based classroom. The following are a few illustrative examples of important classroom exchanges that promote learning but are not targeted by our approach to participation and participation grading.

A type of classroom participation that our analysis ignores is that which occurs when a student answers an instructor's question. The student's answer might change another student's perspective or way of thinking about a topic, but the instructor's question probably wasn't meant to initiate a discussion among the students. We take a similar view of the asking of a question by one student to the instructor or another student, if the person responding simply provides a direct answer. As important as the asking and answering of questions is to learning, we're interested in mechanisms encouraging a different kind of participation.

As a different example, a high-quality discussion doesn't have to be a long discussion. One student can take the initiative to begin a discussion, and we need only have one response for us to begin judging whether the two comments constitute a high-quality discussion. Of course, we would prefer to see a longer discussion, and we will separately report on the changes in the number of replies and participants in our Perusall discussions.

Finally, when talking about authenticity, we have found it important to acknowledge the positive impact that role-playing can have in enriching the learning that takes place in the classroom. We, in fact, use role-playing in some of the exercises in AC221. This should not be viewed in conflict with our general desire for students to bring their authentic selves to the other types of class discussions because this approach encourages perspective-taking amongst students, thereby fostering openness to and respect for others' arguments.

### 5.4 Assessment Rubric

Given the helpful prior work that informed how we would assess authenticity in student comments and the quality of the student discussions on any part of a Perusall reading, we developed and used the rubric in Table 1 to give a numerical rating to each Perusall comment and each Perusall discussion. Each individual comment is evaluated for the characteristics of authenticity (Hadjioannou, 2003) and scored as 0,1, or 2. A comment is scored 0 if it shows no indication of authenticity, 1 if it includes personal opinion or experience, and 2 if the student elaborates on his or her authentic comment by providing related evidence or details.

Each discussion is evaluated based on the characteristics of quality discussions, namely whether students take perspectives by building on their peers' comments within a discussion (Schaeffer et al., 2002). Each discussion is scored as 0,1, or 2. A discussion is scored 0 if it includes only one comment or multiple comments that are "independent and unilateral" (Järvelä, Häkkinen, & Oostendorp, 2003). The discussion is scored as a 1 if students take perspectives in a conversation by building on peers' ideas by responding, acknowledging,  recognizing, and appreciating the value of others' opinions (Bendixen et al., 2003; Järvelä, Häkkinen, & Oostendorp, 2003; Nussbaum et al., 2002). Finally, the discussion receives a 2 if the involved comments further cross-reference one another and incorporate a variety of different perspectives.

*Table 1. Assessment rubric for participation quality measures*

| Metric | Scoring Characteristics | Score: 0 | Score: 1 | Score: 2 |
|---|---|---|---|---|
| **Authenticity of Comments** | Providing personal opinion or experience | Demonstrating negligible or minimal authenticity characteristics | Demonstrating authenticity characteristics | Elaborating on one's own authentic comments |
| **Quality of Discussions** | Taking perspectives by building on previously stated comments | Comments in the discussion fail to build on previous comments | Comments in the discussion build one one another | Comments in the discussion synthesize multiple diverse opinions by cross referencing |

## 5.5. Data Characteristics

We analyzed a total of 2910 comments and 1406 discussions. Spring 2020 data included 2124 comments and 999 initiated discussions and Spring 2021 data included 786 comments and 407 initiated discussions. In Spring 2020, 505 out of 999 discussions included more than one comment and, in Spring 2021, 179 out of 407 discussions included more than one comment. The quality scores were computed for the discussions that include more than one comment. Part of the difference in total comments and total discussions betweens semesters is due to the difference in enrollments between the 2020 and 2021 course offerings. 66 students were enrolled in Spring 2020 and 41 in Spring 2021. We believe, although we cannot prove, that another significant contributor to the difference was that the 2021 students knew that they did not have to comment on every Perusall assignment.

Note that the total number of comments and discussions are derived only from the readings that were assigned in both semesters. This eliminates many alternative explanations that arise when trying to compare the quality and rate of participation on inherently different readings. These counts also exclude instructor comments. The vast majority of instructor comments were not ones that would be considered a part of a discussion. Instead, they were comments about a student's participation. The instructors had informed the students in both years that Perusall was a student space. The instructors would read the conversations that took place in that space and use that information to tailor the subsequent class time (e.g., to skip topics well understood by the students or focus more intently on topics misunderstood by them). The instructors would not be monitoring the space to "correct" student views or make sure that the "right" points were raised in the discussions.

## 5.6. Data Scoring and Reliability Analysis

To maintain the rigor of the analysis, the other author (M.D. Smith) downloaded the data from Perusall, combined the data files, stripped out the identifying features, and randomized the order of presentation to those rating the comments and discussions. He did not assess any of the comments or discussions. Those individuals that scored the comments and discussions did not have access to the semester of origin of the data (e.g., whether a comment belongs to 2020 or 2021).

Once the data was prepared for annotation, the rubric reliability was established. Two raters, one being author (DM) and the other being an external scorer who is an expert in education research and thinking skills, annotated students' comments and discussions separately. Raters scored the authenticity of comments and quality of discussions according to the rubric listed in Table 1 to assess students' participation. The raters annotated approximately 5% of the overall sample (183 comments and 94 discussions). Interrater reliability was achieved with 88% agreement for comments and 90% agreement for discussions. After achieving an acceptable level of interrater agreement on a subsample, one of the authors (DM) annotated the entire sample based on the rubric.

## 5.7. Hypothesis Testing

By using the participation rubric (Table 1), we analyzed the authenticity in comments and quality discussions measures in both Spring 2020 and Spring 2021 semesters. In addition to the quality metrics, we also analyzed several quantity metrics such as the average number of follow-up replies and average number of upvotes in a discussion. In the participation portfolio assessment, students were able to receive feedback from the teaching team, encouraging interactions with peers. Therefore, we expected an increase in the number of replies in Spring 2021. We used a t-test to compare the scores of 2020 and 2021.

## 6. Results
## 6.1. Students' participation in pre-intervention and post-intervention

Students' participation data on Perusall was compared in terms of authenticity of comments and quality of discussions. Table 2 provides the average scores of these metrics as well as the average number of upvotes and number of follow-up replies on Perusall.

*Table 2 Comparison of students' participation in Spring 2020 and Spring 2021 semesters*

| Category of Participation Measures | Spring 2020 | Spring 2021 |
|---|---|---|
| Average authenticity score in comments*** | 0.39 (0.01) | 0.70 (0.02) |
| Average quality score in discussions*** | 0.81 (0.03) | 1.23 (0.05) |
| Average number of upvotes per comment*** | 0.67 (0.03) | 1.24 (0.05) |
| Average number of follow-up replies per discussion | 0.64 (0.01) | 0.65 (0.01) |

*Note.* *** indicates significance at the p <.001 level. Spring 2020 data included a sample of 66 students and Spring 2021 included a sample of 41 students.

The average authenticity score of students' comments in Spring 2021 was statistically greater than that in the Spring 2020 semester ($t$ (1138)= -13.7, $p < .001$). Similarly, the average quality score of discussions in Spring 2021 was also significantly greater than the discussions in the Spring 2020 semester ($t$ (323)= -6.91, $p < .001$). The average number of upvotes per comment was statistically higher in Spring 2020 than that in Spring 2021, $t$ (1283)=-9.34, $p < .001$. The average number of replies per discussion were statistically equivalent to each other, $t$ (277)= -0.8, $p$ =0.37.

Figure 2 and Figure 3 shows the percentage of the authentic comments and the percentage of the quality discussions, respectively. 36 % of comments were classified as authentic in Spring 2020 and this percentage significantly increased to 61% in Spring 2021, ($t$ (1338) = -12.06, $p$ <.001). 62% of discussions in Spring 2020 were classified as quality discussions and this percentage significantly increased to 86% in Spring 2021 ($t$ (442) = 7.13, p<. 001).

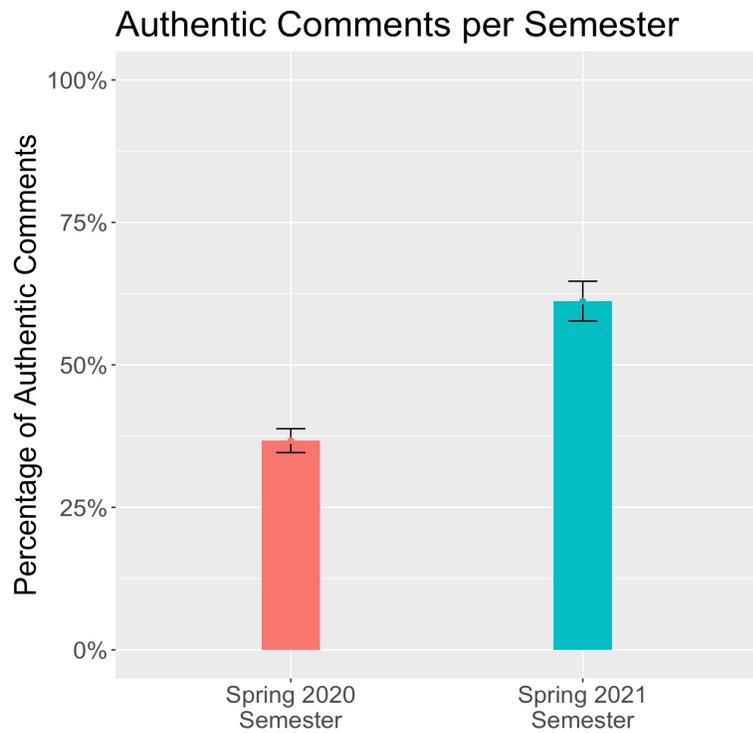

Fig. 2. Number of authentic comments to total number of comments in both semesters

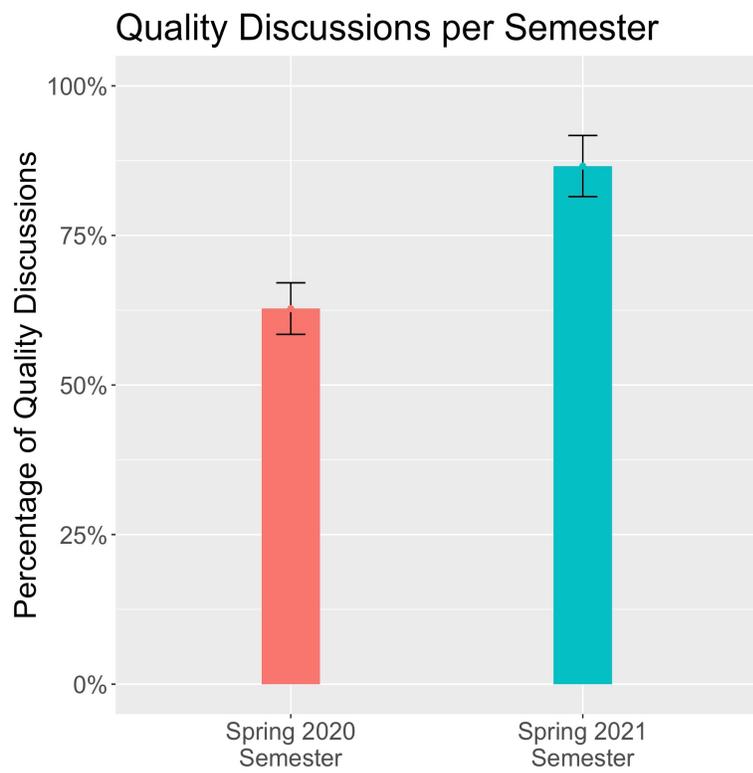

Fig. 3. Number of quality discussions to total number of discussions in both semesters

### 6.2. Students' comments

Below we provide are some representative comments from students about their participation:

*"As someone who did not speak up in many classes prior, I found myself speaking in almost every class,"* said one student in the course evaluations in Spring 2021. Another student emphasized productive discussions that happened with the peers: *"The most interesting part of the course were the class discussions around current events! This class is a fun way to engage with the changes happening right now in the world and to learn from your peers."* The teaching team reported that *"the grading was indeed fun!"* (which many educators would be surprised to hear from an instructor fellow given the burdensome nature of grading).

### 7. Conclusion

This paper has provided a pedagogical framework for designing an assessment tool to foster data science students' motivation for critical thinking. This theoretically-motivated framework, the ARC framework, provides data science instructors practical tools to create an assessment that gives students autonomy in their learning, multiple opportunities to reflect on their learning, and the encouragement to become active participants in a community of learners, all of which are crucial for motivation for learning. Our results suggest that the participation portfolio materially improved students' engagement, motivating them to offer authentic opinions and create higher quality discussions, thus moving toward the learning goals for critical thinking in data science class.

We drew our evidence from two instances of a course in data science (AC221) that was taught in Spring 2020 (pre-pandemic) and Spring 2021 (during the pandemic). One may argue that confounding factors such as the abrupt transition from in-person teaching to remote teaching may have contributed to the significant differences we found.

While other factors may have contributed to these differences, our results, contrary to prior expectations that remote teaching may decrease students' participation, suggest that with effective pedagogical interventions, students' motivation for learning could indeed be improved. Another potential limitation of our study is that we only analyze Perusall data, which is in the form of asynchronous discussions. Our future research will include other forms of classroom discussions and test the effectiveness of the ARC framework in various course elements and in various modalities including in-person and  hybrid learning.

The participation portfolio, as an instance of the ARC framework, provided us the opportunity to ensure that students develop and adjust strategies about their learning, which then eventually increase their motivation for learning. In the process of reporting their participation, students developed reflection skills through which they came to understand potential limitations of their worldviews, assumptions, and biases. In engaging with their peers by challenging them or being challenged by them about their perspectives, students develop critical thinking skills. As they reflect on their interactions with their peers, they also receive feedback from the instructor, which is a crucial component of teaching data science.

All instructors should incorporate timely and meaningful feedback mechanisms into their assessment designs. This is particularly important when teaching students how to tackle real-world problems in data science because the skills required to solve such problems are best learned over time and in conversation with individuals who hold diverse perspectives gained from a lifetime of diverse experiences. These skills cannot be acquired by cramming the night before an exam. As our data science students become aware of the consequences of their thinking through meaningful conversations with their peers, these students come to better understand that there are humans behind data.